# PHOTOPROCESSES FOR THE $^3$H$^4$He AND $^3$He$^4$He CHANNELS IN THE CLUSTER POTENTIAL MODEL

S. B. Dubovichenko, A. V. Dzhazairov-Kakhramanov

Radiative capture in the $^4$He$^3$H and $^4$He$^3$He channels are considered for the $^7$Li and $^7$Be nuclei. The analysis is based on the corresponding two-cluster models. The potentials of cluster interaction include forbidden states and are made to be consistent with the phase shifts of elastic scattering at energies up to 20 MeV. Such an approach is shown to describe the total cross sections for photoprocesses over the entire energy region under consideration.

In our previous papers [1], we obtained intercluster Gaussian potentials that describe phase shifts for elastic scattering in the $^4$He$^2$H, $^4$He$^3$H, $^4$He$^3$He, and $^3$He$^3$He channels at low energies. These potentials include forbidden states (FS), which allows to take into account the Pauli exclusion principle in cluster-cluster interactions. Using these potentials in cluster models, we showed that it is possible to reproduce a some of characteristics of the $^6$Li and $^7$Li nuclei, for which the probability of clustering in the above channels is considered to be comparatively high. Various estimates give a value of 0.6 - 0.8 for the probability of $^4$He$^2$H clustering and a value of ~0.9 for the probability of $^4$He$^3$H clustering. In such systems, states are pure according to Young orbital diagrams [2], and the potentials obtained from phase shifts can be used to describe characteristics of the ground states (GS). In this sense, the situation here is simpler than that in such light systems as N$^2$H, $^2$H$^2$H, N$^3$H, and $^2$H$^3$He, where mixed orbital symmetries are possible in the states with minimum spin. In this case, it is necessary to extract a pure component from interactions obtained using the scattering phase shifts which needed for the analysis of GS characteristics [3].

The analysis of photoprocesses in the cluster model with known potentials is a natural continuation of our previous studies. As before, we assume that a nucleus consists of two structureless fragments and that each fragment is characterized by the properties of the corresponding free particle. The wave function (WF) of the system is not antisymmetrized, but intercluster interactions include FSs. As a result, the WF of relative motion of the clusters has a node structure inside the nucleus. The generalized Levinson theorem is used for phase shifts, and hence, the phase shifts vanish at large energies [2]. Such a one-channel potential model of interactions is rather simple, but it can allow to obtain good results on the basis of potentials whose parameters are determined from experimental phase shifts from the very beginning and are not changed in subsequent calculations.

The total cross sections for photoprocesses were previously calculated in the microscopic potential model [4], which is similar to our model and by the resonating-group method (RGM) [5, 6] for the $^4$He$^2$H, $^4$He$^3$H, and $^4$He$^3$He systems. However, the authors of these studies did not analyze the presence and positions of FSs of the intercluster potentials. Therefore, it is interesting to calculate the cross sections for photoprocesses by using interactions including FSs.

To calculate the capture cross section in the long-wave approximation, we used

the well-known expression [5, 7]

$$\sigma_c(J) = \frac{8\pi}{\hbar^2 q^3} \frac{K^{2J+1}}{(2S_1+1)(2S_2+1)} \frac{\mu}{J[(2J+1)!!]^2} \frac{J+1}{\sum_{m, m_i, m_f}} \left| M_{Jm}(N) \right|^2,$$

$$M_{Jm}(N) = i^J \langle f | H_{Jm}(N) | i \rangle,$$

(1)

$H_{Jm}(E) = Q_{Jm}(L) + Q_{Jm}(S),\ H_{Jm}(M) = W_{Jm}(L) + W_{Jm}(S),$

where, in the cluster model, the operators are given by

$$Q_{Jm}(L) = e\mu^J \left[ \frac{Z_1}{M_1^J} + (-1)^J \frac{Z_2}{M_2^J} \right] R^J Y_{Jm} = A_J R^J Y_{Jm},$$

$$Q_{Jm}(S) = -\frac{e\hbar}{m_0 c} K \left[ \frac{J}{J+1} \right]^{1/2} \left[ \mu_1 \hat{S}_1 \frac{M_2^J}{M^J} + (-1)^J \mu_2 \hat{S}_2 \frac{M_1^J}{M^J} \right] R^J Y_{Jm} =$$

$$= (B_{1J} \hat{S}_1 + B_{2J} \hat{S}_2) R^J Y_{Jm},$$

$$W_{Jm}(L) = i \frac{e\hbar}{m_0 c} \frac{\sqrt{J(2J+1)}}{J+1} \left[ \frac{Z_1}{M_1} \frac{M_2^J}{M^J} + (-1)^{J-1} \frac{Z_2}{M_2} \frac{M_1^J}{M^J} \right] R^{J-1} \hat{L} Y_{Jm}^{J-1} =$$

$$= C_J R^{J-1} \hat{L}\ Y_{Jm}^{J-1},$$

$$W_{Jm}(S) = i \frac{e\hbar}{m_0 c} \sqrt{J(2J+1)} \left[ \mu_1 \hat{S}_1 \frac{M_2^{J-1}}{M^{J-1}} + (-1)^{J-1} \mu_2 \hat{S}_2 \frac{M_1^{J-1}}{M^{J-1}} \right] R^{J-1} Y_{Jm}^{J-1} =$$

$$= (D_{1J} \hat{S}_1 + D_{2J} \hat{S}_2) R^{J-1} Y_{Jm}^{J-1}.$$

(2)

Here, J is the multipole order, q is the wave number of the relative motion of the clusters, μ is the reduced mass, $M_i$, $Z_i$, $S_i$, and L are the masses, charges, spins, and orbital momenta of the interclusters motions, $\mu_i$ are the magnetic moments of the clusters, K is the wave number of the photon, $m_0$ is the nucleon mass, and M is the nucleus mass, R is the intercluster distance. As in [5], the sign of the operator $Q_{Jm}(S)$ is taken to be negative. In what follows, we will take the WFs of the bound states (BS) of the clusters in the form [1]

$$\left| f \right\rangle = \sum_{L_f J_f} R_{L_f J_f} \Phi_{J_f m_f}^{L_f S}, \quad R_{LJ} = \frac{U_{LJ}}{r}.$$

(3)

We write the scattering function in the form [5,7]

$$|i\rangle = \sum_{L_i J_i} i^{L_i} \sqrt{4\pi(2L_i+1)} (L_i\, 0\, Sm_i | J_i m_i) e^{i\delta_{L_i J_i}} R_{L_i J_i} \Phi^{L_i S}_{J_i m_i}$$ (4)

The asymptotic expression for the scattering WF has the form

$$R_{LJ} \sim F_L(qr)\cos(d_{LJ}) + G_L(qr)\sin(d_{LJ}),$$

here $F_L$ and $G_L$ are the Coulomb wave functions, and $\delta_{LJ}$ are the scattering phase shifts. It is rather difficult to calculate numerically the BS WF at distances on the order of 20 - 30 fm. For this reason, the numerical solution for the BS was matched at a distance of 10 - 20 fm with the asymptotic expression for the Whittaker function. The latter has the form

$$R_{LJ} = \frac{\sqrt{2K_0}}{r} C_0\, W_{L\eta}(2K_0 r), \quad W_{L\eta}(2K_0 r) = (2K_0 r)^{-\eta} \exp(-K_0 r),$$

where $\eta$ is the Coulomb parameter, and $K_o$ is the wave number determined by the binding energy. This WF was used to calculate matrix elements. Integration in the matrix elements was carried out numerically with an upper limit of 30 - 40 fm, which ensured high accuracy.

The cross section for the reverse process, photodisintegration, can be obtained from the principle of detailed balance and is written as

$$\sigma_d(J) = \frac{q^2(2S_1+1)(2S_2+1)}{K^2\, 2(2J_0+1)} \sigma_c(J),$$ (5)

where $J_0$ is the GS momentum of the nucleus.

The Gaussian potentials of intercluster interaction used here are given by [1]

$$V(r) = -V_0 \exp(-\alpha r^2) + V_c(r)$$

$$V_c(r) = \begin{cases} \dfrac{Z_1 Z_2}{r} & r > R_c \\ Z_1 Z_2 \left(3 - \dfrac{r^2}{R_c^2}\right) \Big/ 2R_c & r < R_c \end{cases}$$ (6)

The potentials were constructed in such a way that one set of parameters described one partial phase shift in the energy range 20 - 40 MeV. Table 1 displays the parameters of potentials in the $^4\text{He}^3\text{H}$ and $^4\text{He}^3\text{He}$ systems. This potentials differ only by the Coulomb term [1], $\alpha = 0.15747$ fm$^{-2}$ and $R_c = 3.095$ fm.

Table.1. Parameters of potentials in the $^4$He$^3$H and $^4$He$^3$He systems.

| L$_J$ | S | P$_{1/2}$ | P$_{3/2}$ | D$_{3/2}$ | D$_{5/2}$ | F$_{5/2}$ | F$_{7/2}$ |
|---|---|---|---|---|---|---|---|
| V$_0$, MeV | 67.5 | 81.92 | 83.83 | 66.0 | 69.0 | 75.9 | 84.8 |

Table 2 shows the results of these calculations for $^7$Li and $^7$Be, together with the results of RGM calculations [5, 6], results obtained [8] using the potential model with potentials including FSs, and the experimental data reported in [9, 10]. The values obtained for the integral I$_2$ are 13.76 fm$^2$ for the $^3$H$^4$He system and 14.53 fm$^2$ for the $^3$He$^4$He system.

Table 2. Results of calculations for $^7$Li and $^7$Be.

| | Calculation | Experiment | MRG [5] | Calculation [8] |
|---|---|---|---|---|
| $^7$Li | | | | |
| Q, μb | -38.2 | -36.6(3); 40.6(3) | -34.2 ⟩ 41.9 | -37.4 |
| μ/μ$_0$ | 3.383 | 3.2564 | 2.79 ⟩ 3.16 | 3.384 |
| Ω/μ | 2.70 | 2.4(5); 2.9(1) | 3.2 ⟩ 3.6 | 2.48 |
| B(E1), $\mu_0^2$ | 2.45 | 2.48(12) | 1.96 ⟩ 2.17 | 2.45 |
| B(E2) | 7.3 | 7.42(14) ⟩ 8.3(6) | 5.4 ⟩ 11.3 | 7.0 |
| E$_0$, MeV | -2.47 | -2.467 | | |
| E$_1$, MeV | -1.99 | -1.989 | | |
| R$_f$, Fm | 2.40 | 2.39(3) | | |
| R$_m$, Fm | 2.77 | 2.70(15); 2.98(5) | | |
| C$_0$ | 3.9 | | | |
| $^7$Be | | | | |
| Q, μb | -59.3 | | -58.4 | |
| μ/μ$_0$ | -1.532 | | -1.27 | -1.533 |
| Ω/μ, | 2.85 | | 4.71 | |
| B(E1), $\mu_0^2$ | 1.87 | 1.87(25) | 1.58 | 1.87 |
| B(E2) | 17.5 | | 17.0 | |

| $E_0$, MeV | -1.60 | -1.586 | | |
| $E_1$, MeV | -1.14 | -1.157 | | |

Proceeding to the total cross section for photoprocesses, we note that the electric spin operator $Q_{Jm}(S)$ in (2) makes a comparatively small contribution to the total cross section of the process in this systems.

In calculating cross sections, we took into account E1, E2, and M1 transitions, as well as the contribution of the spin part of the electric operator. Let us consider, for example, an E1 transition for which capture can occur either to the GSs with $J = 3/2^-$ and energies -2.47 MeV in $^7$Li and -1.59 MeV in $^7$Be, or to the first excited state with $J = 1/2^-$ and energies -1.99 and -1.16 MeV, respectively. Here, the values of energy are given with respect to the thresholds of the cluster configurations. Capture to the GS of the nucleus occurs from the scattering states S, $D_{3/2}$, and $D_{5/2}$, capture to the first excited state occurs from the S and $D_{3/2}$ states.

Fig.1 illustrates the quality of the description of the phase shifts for $^4$He$^3$H scattering with the above potentials. The $^4$He$^3$H potentials describe the binding energies of the GS and the first excited state, the charge and magnetic radii, the quadrupole and magnetic moments, the probability of the E2 radiative transition $3/2^- \rightarrow 1/2^-$, the octupole moment of $^7$Li (see Table 2), and the elastic and inelastic Coulomb form factors [1].

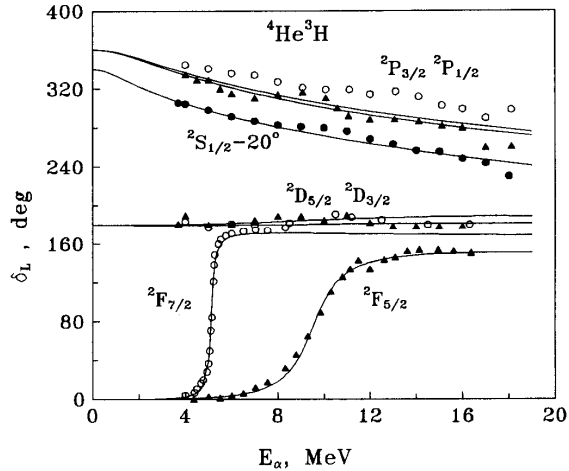

Fig. 1. Phase shifts of elastic $^4$He$^3$H scattering: solid curves is phase shifts calculated with potential (6) and ($\circ$, $\triangle$, and O) experimental data from [11].

Fig.2 shows the calculated cross sections for E1 and M1 capture in the $^4$He$^3$H and $^4$He$^3$He channels, along with the experimental data reported in [7, 12] and the results of calculations for E1 from [7] (dotted curve). The dashed curve in Fig. 2b represents the results of RGM calculations for the E1 transition from [5]. It is clear from Figs. 2 that the cross sections for E2 and M1 processes are smaller than those for E1 processes by 1.5 - 3 orders of magnitude and that they do not make a significant contribution to the total cross section. The peak in the cross section for E2 photodisintegration corresponds to a resonance in the $F_{7/2}$ wave at an energy of

~4.5 MeV (2.16 MeV in the c.m.s.). A similar peak is observed in the cross section for $^4$He$^3$He capture at an energy of 7.0 MeV, this corresponds to the resonance at 2.98 MeV (c.m.s.) relative to the cluster thresholds. The calculated M1 cross sections are in good agreement with the results of RGM calculations [6]. The dash-dotted curve with the label E1' in Fig. 2a shows the cross section associated with the spin term $Q_{Jm}(S)$ of the electric E1 operator. This term in the $^4$He$^3$H system is significantly less than the orbital term $Q_{Jm}(L)$, and its contribution to the total E1 cross section can be disregarded. In these calculations, we used the following values of the magnetic moments of the clusters $\mu_H = 2.9786\mu_0$ and $\mu_{He} = -2.1274\mu_0$.

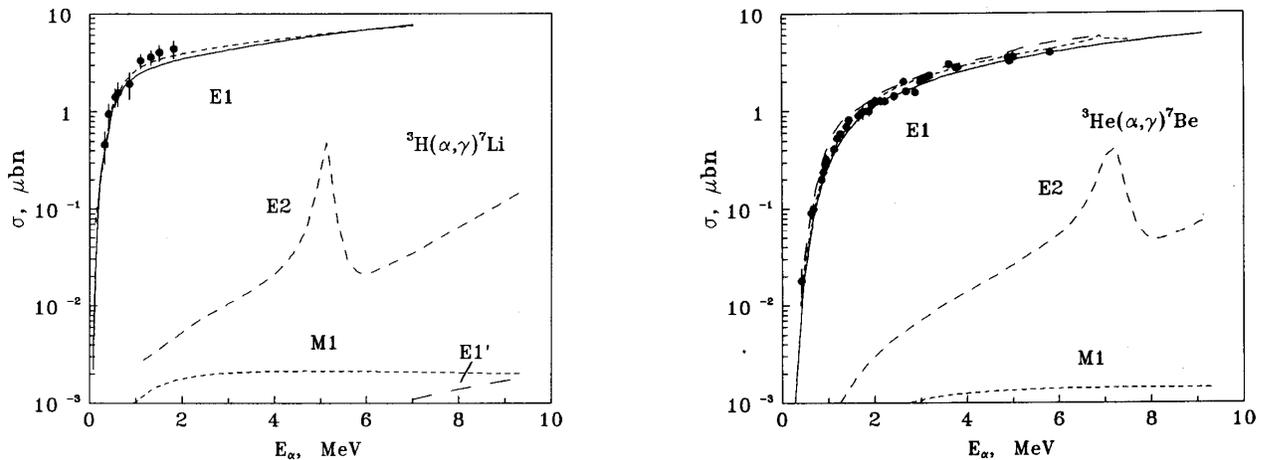

Fig. 2. Total cross sections for the radiative (a) $^4$He$^3$H and (b) $^4$He$^3$He captures with production of $^7$Li and $^7$Be nuclei.

The S factor was also calculated for $^4$He$^3$H capture, it proved to be equal to 0.087 keV b at 20 keV (in contrast to the well-known value 0.064(16) keV b reported in [7] ). It is worth noting that other experimental values are also available, for example, a value of 0.134(20) keV b is presented in [13]. The S factor for $^4$He$^3$He capture at 40 keV equals 0.47 keV b, which is in fairly good agreement with the data from [7, 14], which lie between 0.47 and 0.63 keV b. Calculations in the RGM usually give values from 0.5 to 0.6 keV b [5]. The potential approach yields 0.47(2) keV b [8] or 0.56 keV b [4]. Two sets of experimental data on the S factors are available [7]. They yield S(0) = 0.3(3) and S(0) = 0.61(7) keV b, respectively. The average of these two values - 0.45 keV b is in good agreement with the value obtained here. Figure 3 shows the astrophysical S factors for $^4$He$^3$H and $^4$He$^3$He capture in the energy region up to 3 MeV (these values were obtained from the E1 cross section) in comparison with the results of RGM calculations [5] (dashed curves). The experimental data were taken from [7, 12]. The results of these calculations are in good agreement with experimental data.

It is now clear that the cluster model for interactions involving FSs gives a correct description of static electromagnetic characteristics and total cross sections in the entire energy region accessible to investigation. This also holds true for astrophysical S factors. The large probability of clustering for these nuclei allows suc-cessfully apply the simple one-channel two-cluster model to the description of various nuclear characteristics.

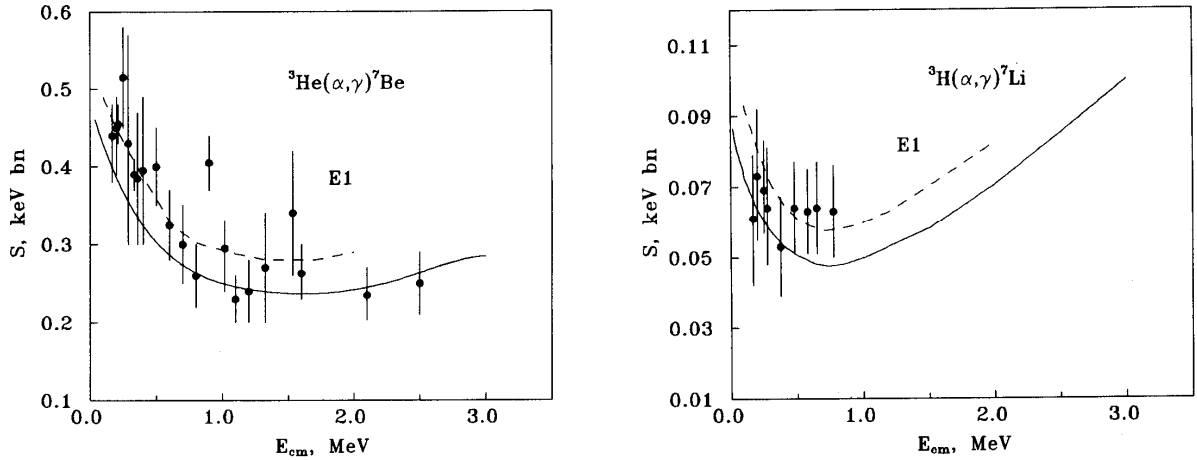

Fig.3. Astrophysical S factors for (a) - $^4$He$^3$H and (b) - $^4$He$^3$He capture.